\documentclass[12pt,preprint]{aastex}
\usepackage{psfig}
\newcommand{\ltsim}{\lower.5ex\hbox{$\; \buildrel < \over \sim \;$}}
\newcommand{\gtsim}{\lower.5ex\hbox{$\; \buildrel > \over \sim \;$}}

\newcommand{\Msun}      {\mbox{$\,M_{\mathord\odot}$}}

\newcommand{\OMC}		{\mbox{${\rm O}-{\rm C}$}}

\newcommand{\Porb}{\mbox{$P_{\rm orb}$}}
\newcommand{\Jorb}{\mbox{$J_{\rm orb}$}}
\newcommand{\Porbdot}{\mbox{$\dot P_{\rm orb}$}}

\newcommand{\exo}{\mbox{EXO\,0748$-$676}}
\newcommand{\xonesix}{\mbox{X1658$-$298}}

\begin{document}

\title{Eclipse Timings of the Low Mass X-ray Binary EXO~0748--676 \\ III.
 An Apparent Orbital Period Glitch Observed with USA and RXTE}

\author{Michael T. Wolff\altaffilmark{1}, Paul Hertz\altaffilmark{2},
Kent S. Wood, Paul S. Ray}
\and
\author{Reba M. Bandyopadhyay\altaffilmark{3,4}}
\affil{E. O. Hulburt Center for Space Research, Naval Research Laboratory,
Washington, DC 20375}

\altaffiltext{1}{E-mail Address: Michael.Wolff@nrl.navy.mil}
\altaffiltext{2}{Now at the Office of Space Science, NASA Headquarters}
\altaffiltext{3}{NRC/NRL Cooperative Research Associate}
\altaffiltext{4}{Now at the Department of Astrophysics, Oxford University}

\begin{abstract}

We present 7 eclipse timings of the low mass X-ray binary \exo\
obtained with the USA experiment during 1999$-$2000 as well as 122
eclipse timings obtained with RXTE during 1996$-$2000. 
According to our analysis, the mean
orbital period has {\it increased} by $\sim$ 8 ms between the
pre-RXTE era (1985$-$1990) and the RXTE/USA era (1996$-$2000). This
corresponds to an orbital period derivative of ${\Porb}/{\Porbdot}
\sim 2 \times 10^7$ years. However, neither a constant orbital period
derivative nor any other simple ephemeris provides an acceptable fit
to the data: individual timings of eclipse centers have residuals of
up to 15 or more seconds away from our derived smooth
ephemerides. When we consider all published eclipse timing data
including those presented here, a model that includes
observational measurement error, cumulative period jitter, and underlying
period evolution is found to be consistent with the timing data. We discuss
several physical mechanisms for LMXB orbital evolution in an effort to
account for the change in orbital period and the observed intrinsic
jitter in the mid-eclipse times.
\end{abstract}

\keywords{binaries: eclipsing, stars: individual: \exo, binaries: X-ray}

\section{Introduction}


The physical process that drives the accretion in low mass X-ray
binary (LMXB) systems is not known with any certainty.  The accretion
may be driven by the loss of orbital angular momentum through
gravitational radiation, by the loss of orbital angular momentum
through magnetic braking, or by the nuclear evolution of the secondary
causing it to overflow its Roche lobe.  The mass transfer from the
donor star to the compact object can be either conservative or
non-conservative.  Models incorporating these physical processes make
specific predictions for the rate at which the system orbital period
(\Porb) changes.  For instance, \Porb\ of an LMXB undergoing
conservative mass transfer at typically observed accretion rates
($10^{-10}$ to $10^{-8}$ $\Msun$ yr$^{-1}$) from a Roche-lobe-filling,
1 $\Msun$ main sequence secondary is expected to decrease with a time
scale of $10^8$ to $10^{10}$ years \citep{rjw82}. Thus, measuring the
orbital period derivative (\Porbdot) should provide a fundamental diagnostic
of the evolution of LMXB systems.  Unfortunately, none of the reported
orbital period derivatives in LMXBs are in agreement with theoretical
expectations (\citealt{tav91}; \citealt{wnp95}).

Six LMXB systems are currently reported to have observed orbital period
derivatives.  Three of these systems (X1820$-$303, \xonesix, and Her
X-1) have apparently negative period derivatives implying that the
orbital separation is shrinking (\citealt{vhd+93}; \citealt{wsb00};
\citealt{dbmk91}).  In each of these systems the orbit period
measurements imply orbital evolution that is proceeding considerably
faster (by 1--2 orders of magnitude) than theoretical predictions for
systems undergoing conservative mass 
transfer \citep[e.g. see the discussion in][]{wnp95}.  
However, for two of these systems
(X1820$-$303 and Her X-1) it is doubtful that the ``standard picture''
of LMXB evolution is applicable.  X1820$-$303 has the shortest orbital
period of any LMXB currently known (11 minutes), the mass donating
secondary is believed to be a white dwarf, and it resides in the
globular cluster NGC~6624, making its orbital dynamics particularly
difficult to understand \citep{vhd+93}.  Her X-1 has a $2.2 \Msun$
companion and is an X-ray pulsar so it is not generally representative
of LMXBs \citep{dbmk91}.  On the other hand, the transient low mass
X-ray binary \xonesix, which resumed its persistent X-ray
emission in 1999 April 2 after a 21-year quiescence, does appear to have a low-mass
donor star, is a prodigious burster, shows kHz quasiperiodic X-ray
oscillations, and shows regular total eclipses in its X-ray light
curve (\citealt{cw84,cw89}; \citealt{ws98}; \citealt{wsf01}; \citealt{wsb00}).  
Thus it appears to be a ``normal'' LMXB.  When the extensive Rossi X-Ray
Timing Explorer (RXTE) observations of \xonesix\ eclipses are combined
with eclipse observations from its previous outburst epoch
\citep{cw89} the results show that the 7.1-hr orbital period of the
system has {\it apparently} decreased by $\sim 11$ milliseconds
\citep{wsb00}. This implies an orbital evolutionary time scale of
$\tau_{orb} \sim |\Porb/\Porbdot| \sim 10^7$ years, and is of a
magnitude similar to other cases of LMXB orbital period evolution
despite the fact that little or no mass transfer is thought to have
occurred during the interval 1978$-$1999.
On the other hand, \citet{ops+01} added two mid-eclipse timings 
based on BeppoSAX observations to the record for \xonesix\ and
concluded that the large \Porbdot\ reported by \citep{wsb00}
may be an artifact of the non-uniform sampling of the
timing data for this source.
Thus, whether or not the orbit in \xonesix\ is 
evolving at an observable rate is still an open question. 

In three other LMXB systems (X1822$-$371, X2127$+$119, and \exo) the
orbital period derivative is apparently positive, implying that the
binary orbital separation is increasing.  The system X2127+119 in the
globular cluster M15 has an observed $\tau_\mathrm{orb} \sim 10^6$ years,
leading \citet{hc98} to conclude that it was in the midst of a
transient super-Eddington accretion episode.  \citet{hn01} studied
X1822-371 and derived an orbital evolution time scale of
$\tau_\mathrm{orb} \sim 3.6 \times 10^6$ years, forcing them to conclude as
well that it was in the midst of a short-lived mass exchange episode.
These two systems, however, only show partial eclipses that must be
fitted with Gaussian profiles in the light curve variations in order
to determine the times of phase 0 for each orbit cycle
(\citealt{hc98}; \citealt{hn01}).  Such a procedure has very large 
error estimates and this makes the observational
determination of their orbit period evolution more uncertain than the
case of either \xonesix\ or \exo.

The LMXB \exo\ is a 3.82 hr eclipsing binary system with a neutron
star primary accreting matter from the Roche lobe-filling, low-mass
main-sequence secondary star UY Vol.  This source exhibits Type 1
X-ray bursts (\citealt{ghpw86}; \citealt{ghpw87}), quasi-regular X-ray
dips (\citealt{smw+92}; \citealt{cbda98}), 1 Hz quasiperiodic
oscillations \citep{hv00} and kHz quasiperiodic oscillations
\citep{hjw+99}, interesting spectral variability \citep{tcs+97}, and
complete periodic X-ray eclipses \citep{pwgg86}.  It is this last
property which makes \exo\ important as there are only three known
LMXBs [\exo, \xonesix, and the recently discovered XTE\,J1710$-$281
\citep{mss02}] showing full eclipses of the compact object X-ray
source.  Eclipsing systems allow the best chance of detecting the
expected orbital period evolution in LMXB systems because the eclipse
transitions should provide accurate fiducial timing marks.  However,
each time new eclipse timings of \exo\ have been made, new conclusions
have been drawn about its orbital period.  These include a constant
\Porb\ \citep{pwgg86}, decreasing \Porb\ \citep{psvc91}, increasing
\Porb\ \citep{adn+92}, sinusoidal \Porb\ \citep{cadn94}, constant
\Porb\ with intrinsic variability \citep[hereafter Paper I]{hwc95}, 
and increasing \Porb\ with intrinsic variability \citep[hereafter
Paper II]{hwc97}.

The usual manner in which orbital period evolution in eclipsing
binaries has been tracked is to observe a small number of eclipses
spaced closely as a group but widely separated in time from previous
measurements (typically by many months or even years). This is done so
that a long time baseline can be established on which to track any
observed changes with a reasonable allocation of observing time. The
measured eclipse times are often fit to a model using a standard \OMC\
(observed $-$ calculated) analysis which is basically a least-squares
fit to a parameterized timing model.  However, \citet{lk93} have argued
that this process can be misleading.  If there is some intrinsic
process that subjects \Porb\ to small random fluctuations around a mean
\Porb\ (which we will refer to as ``intrinsic period jitter''), this
can be misinterpreted as a non-zero \Porbdot, even if the underlying
\Porb\ is constant.  Sparse sampling exacerbates this problem since it
can mask the resulting random walk character of the \OMC\ residuals.
More sophisticated analysis methods (e.g., \citealt{k96}) than the
standard \OMC\ method are required to place confidence limits on the
underlying orbital period derivative in the presence of such intrinsic
period jitter.
 
What is clear is that more than 20 years of monitoring of LMXB orbital
evolution has failed to show any LMXB displaying orbital evolution
that is either consistent with theoretical predictions or that can be
characterized by simple mathematical descriptions.  Beginning with the
launch of RXTE in 1996 we began a program of monitoring \exo\ in an
effort to closely scrutinize its orbital period evolution.  This
program consists of closely grouped observations of up to 6 eclipses
with each group spaced at 2 to 3 month intervals throughout the year.
The launch of the {\it Advanced Research and Global Observation
Satellite} (ARGOS) carrying the Unconventional Stellar Aspect (USA)
experiment further increased the possibilities for eclipse
observations. We discuss the relevant technical aspects of our
observations in
\S 2.

We report here on 7 new eclipse timings determined with USA and 122
new eclipse timings for \exo\ obtained during 1996$-$2000 with RXTE.
Combined with the data reported in Paper II, we now have 4 years of
frequent, accurate (eclipse center timing uncertainty $<$ 1 s) eclipse
timings.  We show in \S 3 that the accurately determined orbital
period during the time interval 1996$-$2000 appears significantly
different than the orbital period observed during the interval
1985$-$1990.  Cycle count has been maintained over the entire 15 years
and we continue to observe significant residuals in eclipse timings
when mid-eclipse times are compared with simple orbital ephemerides
similar to those reported in Paper I.  We significantly update the
maximum likelihood method (MLM) determination of the period evolution
over that of Paper II in \S 4.  MLM solutions for the orbit dynamics
require both orbital period evolution and intrinsic period jitter in
addition to uncorrelated random measurement errors.  In \S 5 we
discuss the observed period evolution and possible physical mechanisms
that may help account for the difference between our results and the
current state of the theoretical models.

\section{Observations}

All of the new observations we present here were made either with the
Proportional Counter Array (PCA) on RXTE or the USA detector on ARGOS.

\subsection{RXTE Observations}

The PCA is an array of five large-area X-ray proportional counters
(Proportional Counter Units or PCUs) with microsecond time resolution
\citep{jsg+96}.  We have conducted a continuing program with the PCA
to observe and time \exo\ eclipses with sub-second timing resolution.
The resolution we obtain is limited primarily by counting statistics
in the X-ray flux and any intrinsic variability in the X-ray emission
from the neutron star.  In the 2-8 keV band the un-eclipsed source
flux is typically $\sim 15 - 25$ counts s$^{-1}$ PCU$^{-1}$, which is
larger than the PCA background in that band ($\sim 3-5$ counts
s$^{-1}$ PCU$^{-1}$).  We restrict our analysis to this energy range
because it provides the best signal--to--noise for our analysis.  We
have conducted 4 to 6 observing campaigns per year since 1996 with the
PCA. During each campaign we observe up to 6 complete eclipses
(starting before ingress and observing until after egress) of \exo\
over roughly a one day period. If RXTE scheduling considerations allow
it, that interval can include a number of consecutive eclipses.

A log of our observing campaigns with RXTE is given in Table \ref{tbl-1}.  For
each eclipse we determine the cycle number ($N$) according to the
cycle numbering system initiated by \citet{psvc91} but with an updated
solution for the mid-eclipse time for cycle 0: $T_0 ({\rm TDB;MJD}) =
46111.07418607$ and $\Porb = 0.159337819$ days.  The results from the
first two campaigns (${\rm N} = 25702 - 26358$) were reported in Paper
II.  For the first three campaigns (${\rm N} = 25702 - 26660$) we
obtained 32 ms resolution binned data with no detector layer
identification (the {\it E\_8US\_32B\_0\_1S} mode of the PCA EDS data
system).  For the remaining campaigns we recorded individual photon
event data (the {\it GoodXenon} EDS mode).

Each RXTE observation is processed in a standard manner using scripts
which call modules of the FTOOLS data analysis package.  Our analysis
technique allows each eclipse to be independently processed in exactly
the same manner.  The script performs the following analysis
functions: (i) The data are filtered for quality and we eliminate
observations where the data are incomplete, where bursts occur at any
time near enough to the eclipse to affect its observed profile, and where the
background contamination is large.  (ii) Where possible (i.e., after
the first three campaigns), we extract only layer 1 data in channels
7--21 which corresponds to an energy range of 2--8 keV. Because we are
interested in timing of eclipses, our results should be independent of
energy-channel boundary considerations.  Thus, we make no adjustment
in our analysis technique across gain-epoch boundaries.  If some of
the PCUs were turned on and/or off during an observation but far from
the eclipse, we extract data only from PCUs that were on during the
entire observation.  (iii) We bin the data in 0.5 s bins (unless
otherwise noted below) and correct the timing markers to the solar
system barycenter.  We use the FTOOL ``faxbary'' which gives us
increased timing resolution over that provided by the barycentering
tool ``fxbary.''  Thus, our data achieve an absolute timing accuracy
of better than 100 $\mu$s (see \citealt{rjm+98}) which is entirely
adequate for our purposes.  (iv) We calculate and subtract the faint
source PCA background model for the epoch of that particular
observation from the binned light curves.  (v) We fit a model eclipse
light curve to each background subtracted, barycenter corrected,
binned light curve.

The model light curve is the standard piecewise linear ramp-and-step
model similar to that used by all investigations which timed \exo\ eclipses
(e.g. \citealt{psvc91}).  The model has seven free parameters: flux
before ingress, duration of ingress, flux during eclipse, duration of
eclipse, time of mid-eclipse, flux after egress, and duration of
egress.  The eclipse fitting routines calculate the best model
parameters using Marquardt's method \citep{ptvf92}.  Our procedure is
to first do a complete fit to all seven parameters in our model and
then with careful refined searching, locate the $\chi^2$ minimum as a
function of mid-eclipse time. The errors in our mid-eclipse times are
estimated by stepping away from the $\chi^2$ minimum in the
mid-eclipse time until we reach the change in this time that
corresponds to an increase in the $\chi^2$ statistic by 1.0 over its
minimum value, refitting only the other six model parameters at each
step.  This procedure yields the experimental 68\% confidence interval
for the single fitted parameter that we are interested in here: the
time of mid-eclipse.

\subsection{USA Observations}

The USA X-ray timing experiment is a two X-ray proportional counter
instrument on the Air Force's ARGOS satellite.  ARGOS was launched on
February 23, 1999, into a sun-synchronous polar orbit from Vandenberg
Air Force Base.  Each USA detector is a thin window proportional
counter with sensitivity between 0.5 and 25 keV and a peak effective
area of 1100 cm$^2$ (\citealt{wfh+99}; \citealt{rwf+01}).  On June 8,
1999, detector 2 of USA failed due to a gas leak.  All of the data
reported here was taken with a single proportional counter.  On
November 17, 2000, the gas system in detector 1 failed as well,
effectively ending the X-ray observational part of the USA mission.

An ongoing USA campaign to observe several \exo\ eclipses a week was
begun on January 9, 2000.  The USA observations of \exo\ provide
photon event data with 32 $\mu$s resolution (USA Science Mode 1).  The
ARGOS polar orbit limits USA observations to exposures which are
shorter than $\sim 1100$ seconds, so observing complete eclipses is
difficult but not impossible. Generally the phase of the ARGOS/USA
orbit allows the observation of either an eclipse ingress or egress,
and if the orbit happens to be very well-phased to the orbit of the
\exo\ system, a complete eclipse. Thus, the USA observation database 
has numerous short observations that include either an ingress or an
egress.  The observations we present here are 7 observations which
included full eclipses.  In Table \ref{tbl-2} we give a list of USA observations
of full eclipses included in our analysis. The USA orbital constraints
also make it impossible to observe consecutive ($\Delta{\rm N} = 1$)
\exo\ eclipses.

For each observation, we have binned the data into 1.0 s bins (unless
otherwise noted below) and corrected the timing markers to the solar
system barycenter including all relevant solar system corrections from
the DE200 ephemeris. The ARGOS satellite positional accuracy is
sufficient to provide us with absolute timing accuracy to better than
300 $\mu$s, which is again good enough for our purposes
\citep{rwb+00}.  Once the usable sections of data are identified we
can proceed for the USA observations in the same manner as for the
RXTE observations.  We fit the same type of seven-parameter linear
ramp-and-step model as in the RXTE case to the full eclipses observed
by USA.

\subsection{Other Published Data}

Numerous other investigations have been made of the orbital ephemeris
of \exo\ and we have incorporated those data in several places in our
analysis.  We have taken 30 EXOSAT mid-eclipse times and one GINGA
mid-eclipse time from \citet{psvc91}, eight GINGA mid-eclipse times
from \citet{adn+92}, four ASCA mid-eclipse times from \citet{cadn94},
and one ROSAT timing from \citet{hlwc94}. This yields an additional 44
mid-eclipse timings dating back to February 1985.  These data are listed in
Table \ref{tbl-3} for completeness.  Where appropriate, we have
adjusted the given UTC-based barycenter corrected times to the TT time
system by adding the correct number of leap seconds and adding the
32.184 second difference between the TAI and TT time systems. 
In most cases we have adopted the errors stated in those papers except that we
assign a uniform error of 1.5 seconds to the \citealt{cadn94}
points. Thus, when we refer to ``All Data'' all the times incorporated
in our analysis are on the same time system and thus directly
comparable.

\section{Orbital Ephemerides}

\begin{figure}
\centerline{\includegraphics[height=6.0in,angle=270.0]{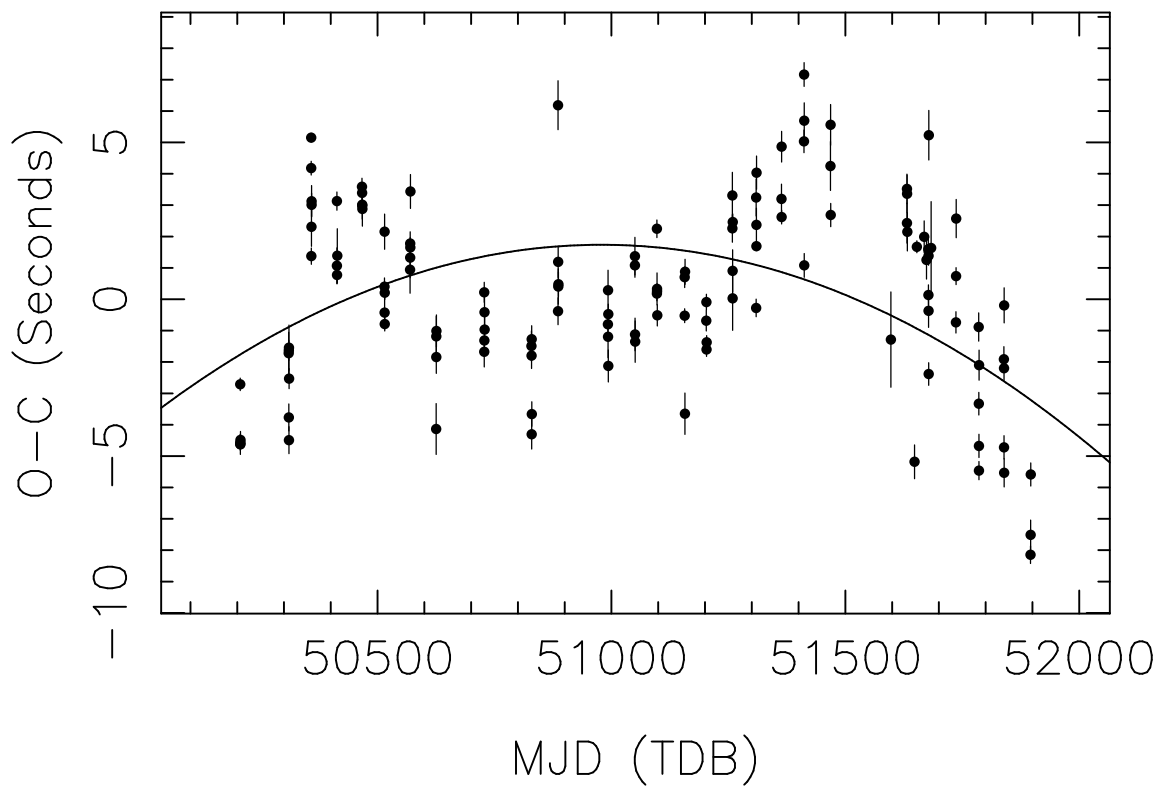}}
\caption{Mid-eclipse timing residuals for USA and RXTE observed eclipses of
\exo\ during 1996$-$2000.  The residual of the observed mid-eclipse
time from a constant period of 0.15933781910 d and epoch of phase zero
46111.07418607 (MJD;TDB) is plotted as a function of barycenter
corrected observation date. The curved line is the constant period
derivative fit to the RXTE and USA data from Table \ref{tbl-4}.  The residuals
appear grouped both above and below the ${\rm O}-{\rm C} = 0 $
baseline.  The measurement errors are significantly less than the
systematic wandering in the mid-eclipse times as was found in Paper II.
\label{fig-omc1}
}
\end{figure}

In Figure \ref{fig-omc1} we show a traditional \OMC\ diagram (observed
minus calculated mid-eclipse time) for the USA and RXTE \exo\ eclipse
timings from Tables \ref{tbl-1} and \ref{tbl-2} where the calculated times are the
best-fit constant period ephemeris (given in Table \ref{tbl-4}).  The error bars
in the figure represent the {\it measurement errors} for each
observation as determined in \S 2.1.  The fit is formally very poor,
with a reduced $\chi^2$ statistic of 84.8 per degree of freedom (dof)
for 127 dof.  The rms scatter in the residuals is 2.94 seconds which
is much larger than the rms standard measurement uncertainty of 0.47
seconds.  There is clearly significant variability in the measured
mid-eclipse times about the ephemeris. The variation in the \OMC\
values can not be explained by simple measurement error.  When the USA
and RXTE data (1996$-$2000) are fitted with a non-zero \Porbdot\ model
the best fit shows a {\it negative} (\Porb\ decreasing) period
derivative.  This fit is also very poor (see the solid curve in
Figure~\ref{fig-omc1}) with $\chi^2/{\rm dof} = 68.0$, almost as large
as the constant \Porb\ case.  The rms scatter of the residuals is now
2.78 s.

\begin{figure}
\centerline{\includegraphics[height=6.0in,angle=270.0]{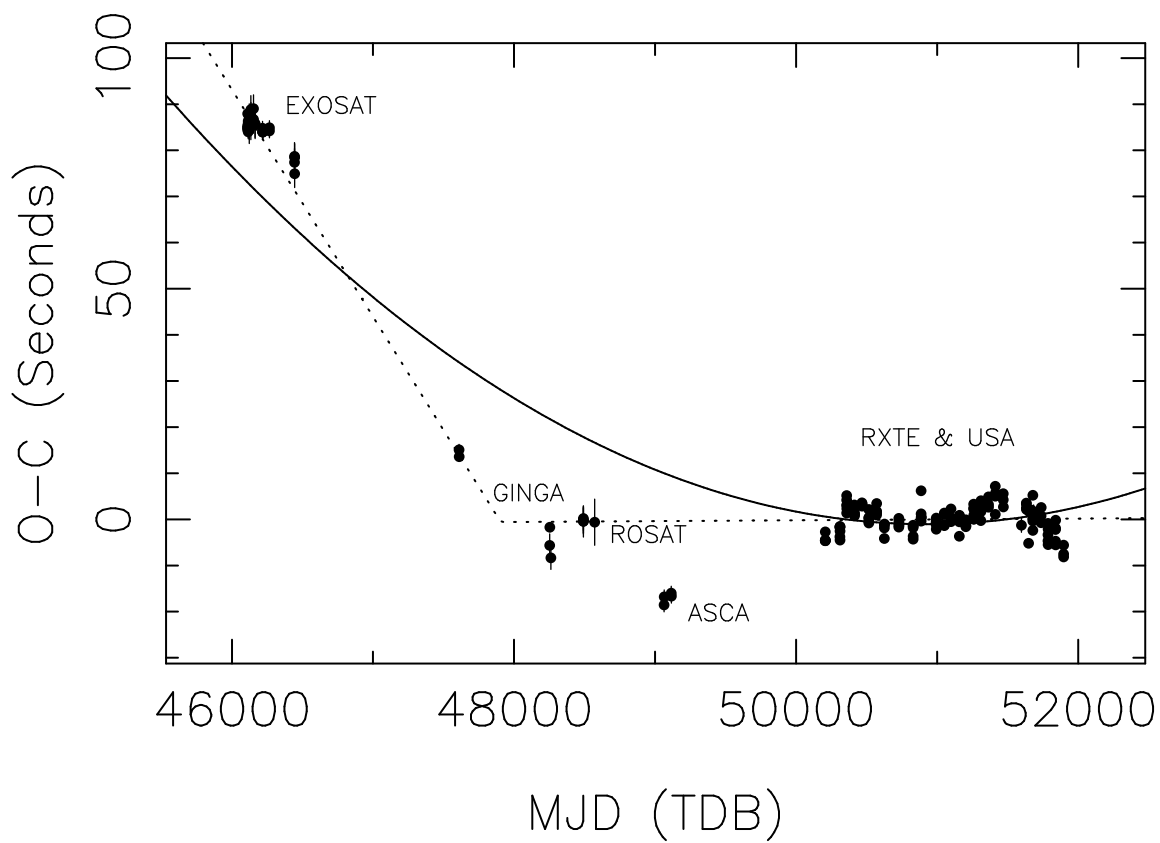}}
\caption{The mid-eclipse timing residuals for observed eclipses of \exo\ during
1985$-$2000 both from the present study and those available in the
literature.  The residual of the observed mid-eclipse time from the
same constant period model as in Figure \ref{fig-omc1} is plotted as a
function of barycenter corrected observation date. The curved solid
line is the constant period derivative solution to all the data from
Table \ref{tbl-4}.  The dotted line is the broken constant period solution from
Table \ref{tbl-4}. No simple linear or quadratic ephemeris fits all the data
points.
\label{fig-omc2}
}
\end{figure}

In Figure~\ref{fig-omc2} we show the \OMC\ residuals to a same
constant period model for {\em all} published mid-eclipse timings of
\exo. Obviously, a constant period model is unacceptable.  The EXOSAT
points have \OMC\ residuals of 75--90 seconds relative to the constant
period ephemeris.  A constant \Porbdot\ model (whose parameters are in
Table \ref{tbl-4}) is also a poor fit to the data with a reduced $\chi^2 $ of
$121.1$ with 170 dof.  The residuals in Figure~\ref{fig-omc2} also might be
interpreted as evidence for a sudden period change around 1990.  Thus,
we have also fitted a two--period model to the data in which we
constrain the phase to be constant across the instantaneous period
change but let the cycle of the period change be a free parameter.
The best fit two--period model is given as the last model in Table \ref{tbl-4}.
In that model, $P_\mathrm{orb,1}$ is tightly constrained by the large
number and high precision of the RXTE observations.  The change in
period is $P_\mathrm{orb,1} - P_\mathrm{orb,0} = 7.87
\pm 0.06$ ms.  However, even this model is a statistically poor fit to
the timing residuals with a reduced $\chi^2/{\rm dof} = 67.3$ for 169
dof.  Furthermore, as we show below, it is likely that 
the ``sudden'' period change is simply an
artifact of the much sparser sampling in the older data.

\section{Observed Intrinsic Period Jitter and Eclipse Profile Changes}

\subsection{Maximum Likelihood Analysis}

During the epoch covered by the RXTE and USA observations, 1996--2000,
there are significant residuals in the mid-eclipse timings about a
constant period.  These residuals can appear on time scales as short
as a single orbital cycle: the observed change in orbit period as
determined across consecutive mid-eclipse timing fits differs by over 11
seconds, ranging from $0.02 \pm 0.29$ for cycles 25704--25706 to $11.38
\pm 0.53$ s for cycles 34938--34940.
Furthermore, just as was found in Paper II, the \OMC\ residuals in
Figure~\ref{fig-omc1} and Figure~\ref{fig-omc2} are strongly
correlated.  In Paper II we pointed out that apparent changes in
orbital period can be caused by cumulative intrinsic jitter in a
system with constant mean orbital period.  We make a distinction here
between ``measurement error'' and ``intrinsic period jitter'' in the
mid-eclipse timings.  Measurement error is the random,
uncorrelated error associated with our measurement of the individual
mid-eclipse times and is dominated by counting statistics and short
time-scale ($\sim$seconds) fluctuations of the source flux that
degrade the accuracy of our measured eclipse profiles.  On the other
hand, by intrinsic jitter we mean a cumulative process where the
orbital period for any particular cycle suffers small random (zero
mean) fluctuations around the true underlying orbital period. 
If the underlying model used to compute the predicted eclipse times is
correct, then the \OMC\ residuals can be represented as  
\begin{equation}
(O-C)_j = \sum_{i=1}^{N_j} \epsilon_i + e_j,
\label{eqn-randomwalk}
\end{equation}
where $\epsilon_i$ is a random, zero-mean, fluctuation in the length
of orbit period $i$ and $e_j$ is the measurement error in the $j$th
mid-eclipse time \citep{k96}.  The cumulative nature of the
$\epsilon_i$ causes the systematic wandering of the mid-eclipse
residuals apparent in Figure~\ref{fig-omc1}. 
In other words, the mid-eclipse times are doing a random walk 
about the times that would be predicted from 
the ``true'' orbital period.

We have used Koen's (1996) maximum likelihood method (MLM) to estimate
the parameters of a model for the orbital evolution that
simultaneously includes an orbital period, an orbital period
derivative, intrinsic period jitter, and random measurement error.  In
Paper II we applied a similar method to a small subset of these
data. In the present analysis we include all eclipse timings given in
Tables \ref{tbl-1} and \ref{tbl-2} as well as all previously published
eclipse timings given in Table \ref{tbl-3} for a total of 173 mid-eclipse 
times.  In the MLM method a likelihood ratio statistic is used to
determine how important a certain parameter is in fitting these data.
The observed mid-eclipse timings are represented by $T_0$, $T_1$,
$T_2$, ..., $T_J$, corresponding to cycle numbers $N_0$, $N_1$, $N_2$,
..., $N_J$, where $J$ is the number of observed eclipse centers.  The
time intervals between successive eclipse centers are given by $y_j =
T_j - T_{j-1}$, with each interval accounting for $n_j = N_j -
N_{j-1}$ cycles.  If a linear form is assumed for the period evolution
then it is easy to show that the $y_j$ can be written as
\begin{equation}
y_j = n_j P_{orb} + m_j \Delta + \sum_{i=N_{j-1}+1}^{N_j} \epsilon_i
 + e_j - e_{j-1},
\end{equation}
\noindent{where} $m_j = n_j ( 2 N_j - n_j + 1 ) / 2$, 
$\Delta = \Porb \Porbdot$, $e_j$ is the observational measurement
error for cycle $j$, and $\epsilon_j$ is the intrinsic jitter in the
mid-eclipse timing for cycle $j$.  The MLM method consists of solving
a linear system of J equations for the maximum likelihood estimates
for the orbit period \Porb, the orbit period derivative $\Delta$, the
measurement error variance $\sigma_{e}^2$, and $q \equiv
\sigma_{\epsilon}^2 / \sigma_{e}^2$, where $\sigma_{\epsilon}^2$ is
the variance of the intrinsic jitter.  For a particular value of $q$
one can define the logarithm of the likelihood function by
\begin{equation}
\ln L(H_i) = - {1 \over 2} [ J {\rm ln} ( 2 \pi ) + 
J {\rm ln} ( \sigma_e^2 ) + {\rm ln} | \Sigma_{*} | + J], 
\end{equation}
\noindent{where}\ the $L(H_i)$ represents the
likelihood of the particular hypothesis ($i$) one is testing, and
$\Sigma_{*}$ is the covariance matrix for the $y$-values divided by
the measurement error variance (see Koen 1996 for details).  The
``full'' model consists of a representation of the data in which each
of the four parameters, $\sigma_{e}^2$, $\sigma_{\epsilon}^2$, \Porb,
and $\Delta$, take on non-zero values and is labeled hypothesis $H_0$.
This model is determined by iterating on the value of $q$ until $
\ln L(H_0)$ attains its maximum value.  Once the full model is found it
can be compared with model solutions for which either the orbital
period derivative is zero ($\Delta = 0$; {\rm hypothesis} $H_1$), or
the intrinsic jitter ($\sigma_{\epsilon}^2$) is zero ($q = 0$; {\rm
hypothesis} $H_2$).  The likelihood ratio statistic
\begin{equation}
\lambda_k \equiv 2 [ \ln L(H_0) - 
\ln L(H_k) ] \,\,\,\, \mathrm{for} \,\, k = 1,2 \,,
\label{eqn-likelihood}
\end{equation}
\noindent{then} gives a measure of the likelihood of the full model 
relative to one of the null hypotheses.

In Table \ref{tbl-5} we give results for two cases: (1) all
data including the USA and RXTE timings, and, (2) the USA and RXTE
data only.  
For case (1) the preferred model is a model that includes
both non-zero intrinsic period jitter ($\sigma_{\epsilon} > 0$) and
period evolution (\Porbdot\ $\neq 0$).  
A model for the full data set that has no period evolution is rejected 
relative to the full model at the 99.86\% ($e^{-7.16}$) confidence 
level ($\lambda_1 = 10.16$) and a
model with no intrinsic jitter is rejected when compared to the full
model at the $\gg 99.999$\% ($e^{-207.8}$) confidence level
($\lambda_2 = 407.7$).  
For case (2), where only the 1996$-$2000 USA
and RXTE data are considered, the results are not so clear.  
A model with no period evolution ($\Delta = 0$) but non-zero intrinsic jitter
is only rejected at the relatively weak 90.3\% ($e^{-2.80}$)
confidence level compared to the full model ($\lambda_1 =
2.76$). 
Again, however, a model with no intrinsic jitter is strongly
rejected at the $\gg 99.999$\% ($e^{-41.6}$) confidence level
($\lambda_2 = 76.9$).  
Furthermore, the derived \Porbdot\ from the
full model for the RXTE and USA case is negative just as it was for
the simple $\chi^2$ fitting in the previous section and this is
contrary to the results for \Porbdot\ when all eclipse data are
included. 
These results indicate that during the RXTE and USA era,
because of the apparent intrinsic period jitter, the mid-eclipse
timing data are insufficient to clearly determine the true long-term
behavior of the orbit period in \exo.  The intrinsic jitter in the
eclipse timings contributes significantly to the apparent variation in
the orbit period over the past 4 years. For this system at least, the
intrinsic jitter appears to mask the underlying orbit period behavior
over multi-year time scales.

\begin{figure}
\centerline{\includegraphics[height=6.0in,angle=270.0]{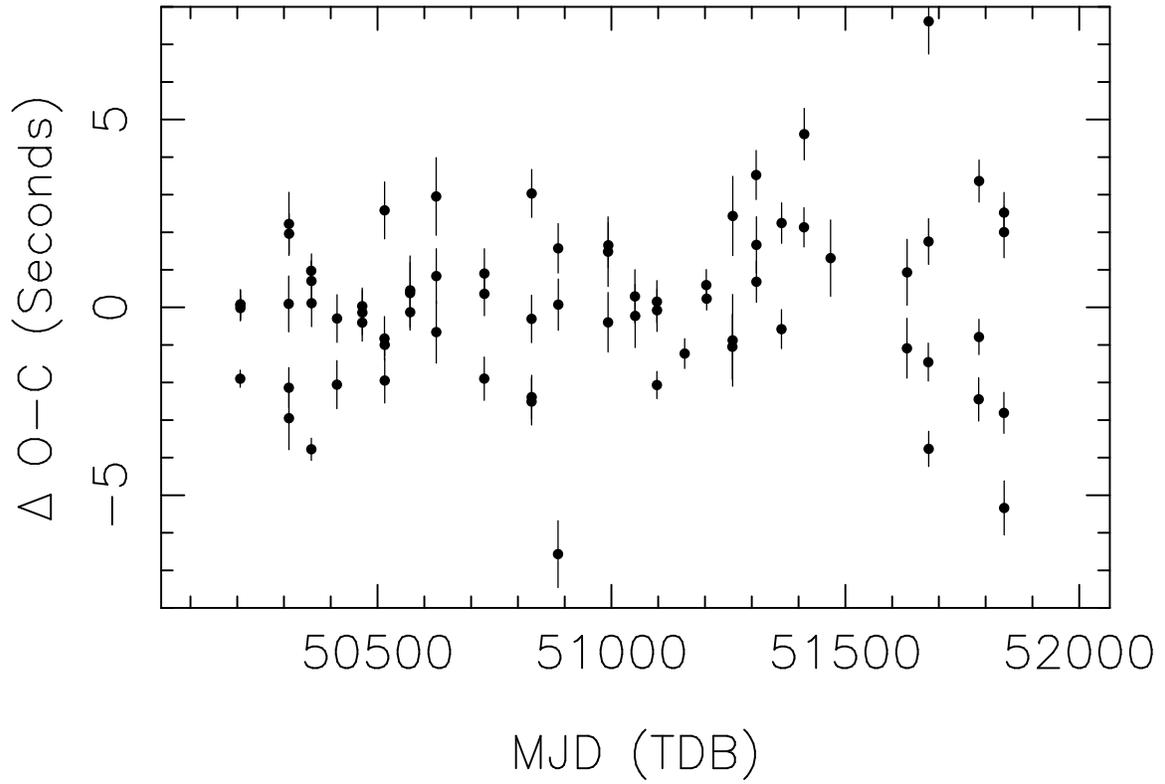}}
\caption{The difference in the \OMC\ residuals for the case of consecutive
observed eclipse cycles.  We plot here ${(\OMC})_n - ({\OMC})_{n-1}$
when the observed eclipse cycle number $n$ increases by 1.
\label{fig-omcdiff}
}
\end{figure}

There is one important caveat that must be mentioned in evaluating the
MLM analysis above, however.  In deriving Equation
(\ref{eqn-likelihood}) the assumption is made that both the
measurement error and the intrinsic jitter are independent and
normally distributed random variables.  For the measurement error this
is a reasonable assumption but for the intrinsic jitter the situation
is less clear.  In Figure~\ref{fig-omcdiff} we see that the \OMC\
residuals for 73 pairs of observed consecutive eclipses during the
RXTE and USA era are scattered around a mean difference of 0.061
seconds with a standard deviation is 2.21 seconds.  The amplitude of
the point-to-point variations in this figure is determined by the sum
of the measurement error and the intrinsic jitter.  This level of
variation is close to the sum of the measurement error and intrinsic
jitter from either the MLM full model or the no period evolution
model in the case of the RXTE and USA data only (see Table \ref{tbl-5}).  Thus,
based on these data, we can not determine if the intrinsic jitter is
truly random and independent because any variations would be swamped
by the measurement error in this figure unless they were very large.
But we can observe that if the period varied in some systematic way
during the RXTE/USA era due to large variations in the intrinsic
jitter then a shift away from a mean of zero with time would be
apparent in the figure but no shift is observed.  Furthermore, if the
magnitude in the intrinsic jitter changed dramatically during the
RXTE/USA era then the amplitude of the variations in the figure would
reflect this but again no such effect is observed. An independent
check on the properties of the intrinsic jitter must await the
observation of substantially more eclipses than we present here.

\begin{figure}
\centerline{\includegraphics[height=6.0in,angle=270.0]{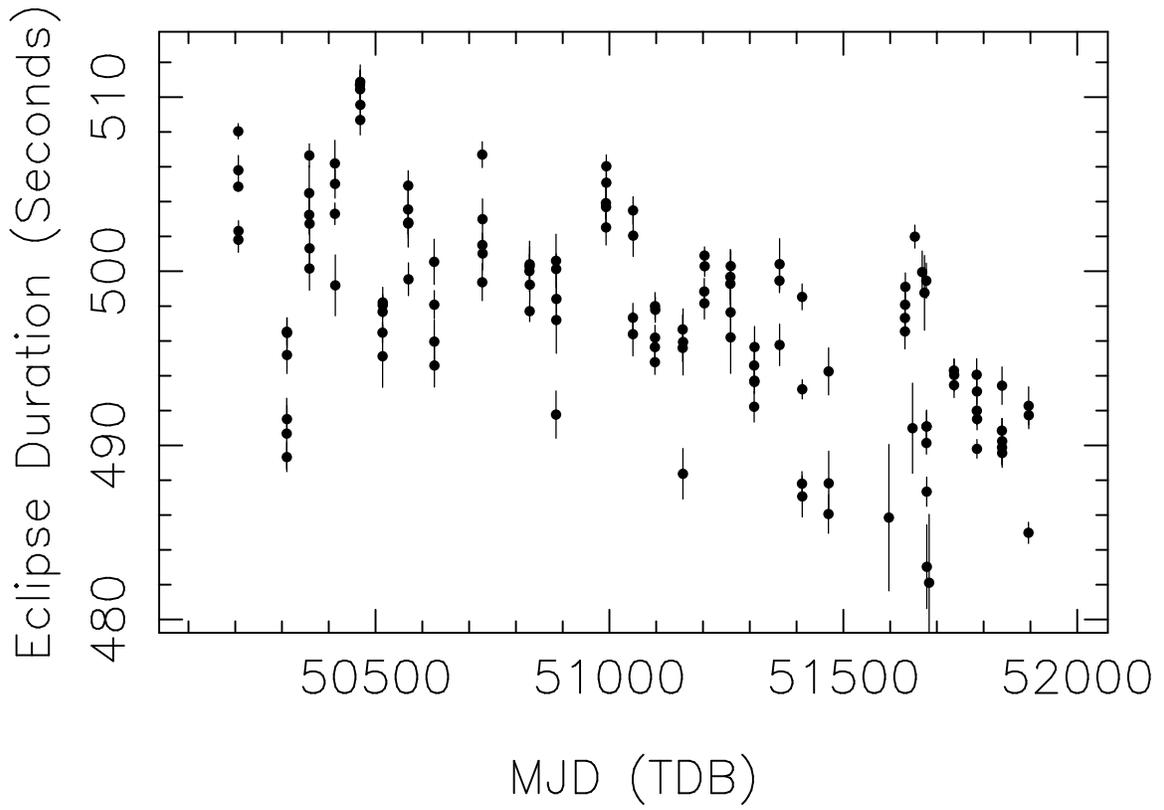}}
\caption{Eclipse duration of \exo\ during 1996$-$2000 as a function of
observation date.
\label{fig-duration}
}
\end{figure}

\begin{figure}
\centerline{\includegraphics[height=6.0in,angle=270.0]{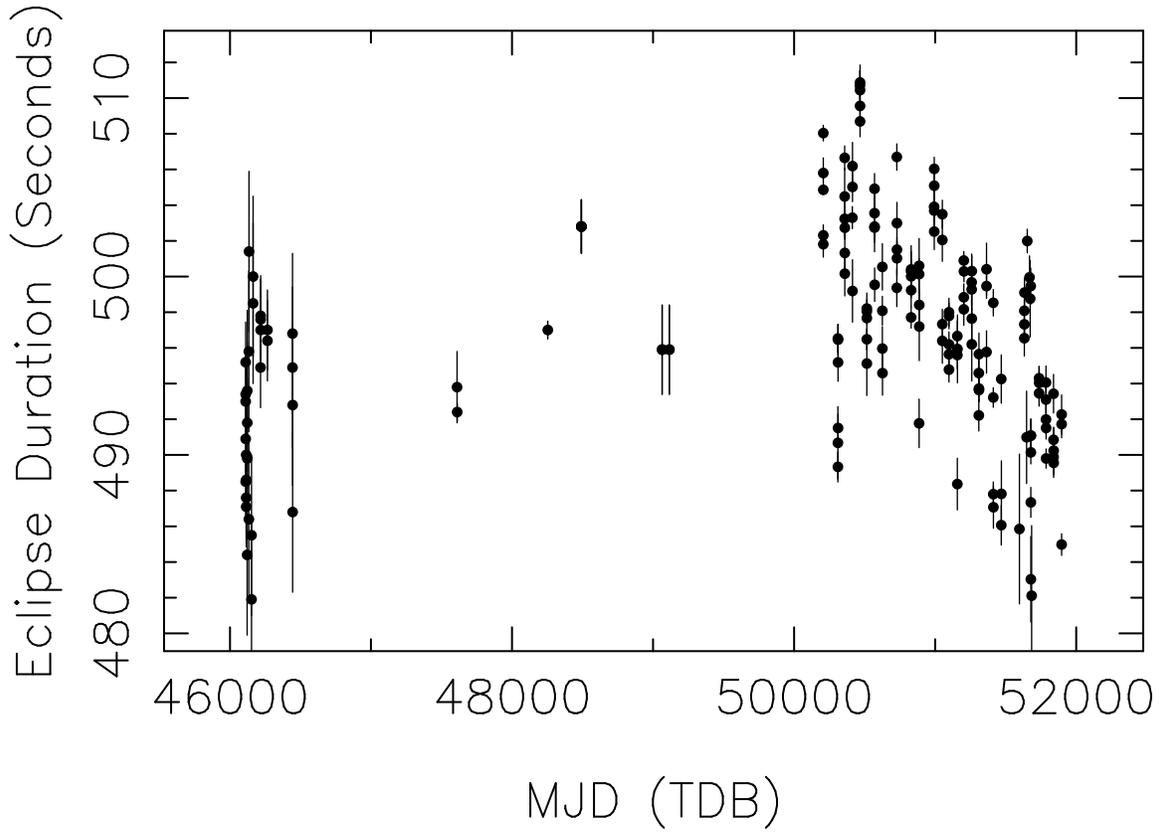}}
\caption{Eclipse duration of \exo\ during 1985$-$2000 as a function of
observation date.  Note that (i) the average eclipse duration was
shorter in 1985$-$1986 than it was in 1996$-$1999, and (ii) the
eclipse duration is variable during any epoch.
\label{fig-duration2}
}
\end{figure}

Finally, we note that the eclipse profiles can be quite variable. 
We define the ingress and egress durations as the time it takes for 
the observed count rate to go from the pre-transition count rate to 
the post-transition rate, and the eclipse duration as the time
from the end of ingress to the beginning of egress. 
With these definitions we find that the
RXTE and USA observed eclipse durations vary between 482 s and 511 s
during 1996$-$2000 (Figure~\ref{fig-duration}) with mean $497.5 \pm 6.0$ s, 
while the average duration of the EXOSAT-observed eclipses in 1985$-$1986 
was $492.5 \pm 5.0$ s (\citealt{psvc91}; Figure \ref{fig-duration2}).  
The duration of ingress and egress varies as well with observed ranges 
between 1.5 s and 40 s for EXOSAT and RXTE observed 
eclipses (Figure \ref{fig-transitions}).  
The EXOSAT observed eclipse transitions averaged 
about 11.7 s, and the RXTE observed eclipse transitions 
averaged about 7.5 s in duration. 
There appears to be a slow downward trend in the eclipse duration during 
the RXTE/USA era apparent in Figure 4. 
We note, however, that the eclipse durations appear to be uncorrelated 
with other system eclipse parameters (e.g., the durations of either ingress
or egress transitions) during this era. 
This variability and the intrinsic variability of the X-ray source being
eclipsed contribute to the uncorrelated random error and explain why
the measured $\sigma_e$ is larger than the measurement errors derived
from the eclipse fits.

\subsection{Monte Carlo Illustration}

To illustrate how the a random walk and sparse sampling can interact
to exhibit the behavior we have seen in the \exo\ residuals, we have
made a simple Monte Carlo simulation which implements Equation
\ref{eqn-randomwalk} for a simulated eclipsing system.  
We assume that the system has no underlying
orbital period derivative and that observed mid-eclipse times are subject 
to both measurement error and intrinsic period jitter.  
In Figure \ref{fig-monte} we show both the full
simulated dataset and one that has been subjected to the same sampling
function as the actual \exo\ measurements. Comparing this to Figure
\ref{fig-omc2} demonstrates how such a simple model can replicate
the behavior we observe in a real system.  Looking at the sparsely sampled 
data one could easily see, or fit, spurious period derivatives, sudden period
changes, or other phenomena that are caused by the random walk and
are not present in the underlying period evolution (to which the random
walk will always return over long time scales).  
Furthermore, when we take the results for the eclipse timings from our Monte Carlo 
simulations, with a sampling function that represents the real sampling of the
\exo\ eclipses, and apply the MLM analysis technique to these data we recover 
the approximate intrinsic jitter magnitude and the estimated measurement 
error magnitude we used as inputs to the Monte Carlo model. 
This demonstrates that the MLM results are a valid representation of the timing data.  
Since there was no \Porbdot\ in the Monte Carlo simulations, the MLM technique
should not indicate a preference for a non-zero \Porbdot\ and this is indeed 
the case.
In the case of the real \exo\ data the MLM model indicates that the 
random walk is unlikely to be able to fully
account for the $\sim 90$ second residuals observed in the EXOSAT era, and thus
a model with some period evolution plus the intrinsic jitter is preferred.

\section{Discussion}

Based on our analysis above we conclude that the \exo\ orbital period has 
increased by $\sim8$ ms over the past 16 years.  
This measured \Porbdot\ implies that the two stellar components 
of \exo\ are moving away from each other instead of toward each other 
as our current theoretical understanding of LMXB orbit evolution 
indicates they should (\citealt{rjw82};
\citealt{tav91}; \citealt{vv95}).
Furthermore, the intrinsic period jitter that we conclude causes the 
measured mid-eclipse times to do a random walk in the \OMC\ diagram 
is certainly not expected from previous theoretical work, but we are 
forced, by the data, to conclude that it is present at the level 
of $\sim 0.1$ s per cycle.  We now turn to some possible explanations 
for the behavior we observe in \exo.

\subsection{A Positive Orbital Period Derivative?}

Mass capture by the compact object in \exo\ can lead to a positive
\Porbdot\ as we show with a simple model.  If the stellar component
spin angular momentum is negligible and the orbital eccentricity 
is zero, the total system angular momentum
consists of only orbital angular momentum which we can write as
\begin{equation}
\Jorb = \sqrt{\frac{GaM_1^2M_2^2}{M_1+M_2}} \,\,\, ,
\label{eqn-j}
\end{equation}
where $M_1$ is the mass of the primary star (the compact object)
that accretes and $M_2$ is the mass of the secondary mass-losing star,
$a$ is the orbital separation and G is the gravitational constant.
Differentiating equation (\ref{eqn-j}), using Kepler's Law, and assuming conservative 
mass transfer gives  
\begin{equation}
{\Porbdot \over \Porb} = 
\frac{3}{\Jorb}\left({\frac{\partial \Jorb}{\partial t}}\right)_{GR} -
\frac{3 \dot{M_2}}{M_2} \left[ 1 - \frac{M_2}{M_1} \right],
\label{eqn-pdot}
\end{equation}
where the first term represents the loss of system angular momentum to
gravitational radiation and $\dot M_2$ is the mass loss rate from the
secondary star.  We ignore magnetic braking here since the interior of
the low mass secondary is likely to be completely convective and
magnetic braking under such circumstances is believed to be suppressed
\citep{vv95}.  In Figure \ref{fig-pdot} we show possible solutions of 
Equation (\ref{eqn-pdot}) for systems with the range of parameters
thought to be relevant to \exo\ \citep{pwgg86}.  In solving this
equation we have ignored considerations of Roche lobe radii and
stellar evolution and only constrained the accretion rate to the range
found by \citet{ghpw86} in their analysis of X-ray bursts from the
\exo\ system.  The observed period derivative ($\Porbdot \sim 1.9
\times 10^{-11}$) requires a secondary mass less than that found by
\citet{pwgg86} in their analysis of the EXOSAT observations ($M_2 \sim
0.45 \Msun$) but still within their allowed secondary mass range.
However, this mass estimate was based on the assumption of the
secondary star filling its Roche lobe and Equation (\ref{eqn-pdot}) is
derived without imposing this requirement.  Thus, as long as we do not
require any specific orbital geometry for the \exo\ system a positive
\Porbdot\ is allowed by simple theoretical considerations.

Negative values of \Porbdot\ are the result of imposing on the LMXB
models the additional constraints of Roche lobe geometry, stellar
evolution of the binary components, and magnetic braking (e.g.,
\citealt{prp02} and references therein).  Such models require that
gravitational radiative angular momentum losses and magnetic braking
conspire with the nuclear evolution of the secondary to slowly shrink
the orbital separation and the secondary's Roche lobe resulting in
mass transfer through the inner Lagrangian point.  The negative
\Porbdot\ trend is maintained until the system attains an age near a
Hubble time when a minimum period is reached and \Porbdot\ becomes
positive as the orbital evolution is driven solely by gravitational
radiation and the secondary is only a low mass degenerate helium star.
A positive \Porbdot\ before this final evolutionary epoch implies an
expansion of the orbit radius and thus the Roche lobe radius, contrary
to theoretical expectations.  If one wishes to retain a gravitational
radiation driven mass transfer model for the \exo\ system, the detailed
models suggest you can not simultaneously reproduce the derived
secondary mass and the observed orbit period while also meeting the
observational requirement of a positive \Porbdot.  Therefore, some additional
physical mechanism other than the nuclear evolution of the secondary,
angular momentum loss via gravitational radiation, or magnetic braking
may be at work in the \exo\ system.
However, if the secondary has a radius significantly less than
its Roche lobe radius then some other mechanism besides Roche lobe
overflow must be found to facilitate the observed mass transfer.

\citet{hnr01} showed that as the secondary star in cataclysmic 
variables loses mass and evolves its outer envelope can be severely
out of thermal equilibrium.  This lack of equilibrium can result in
the star becoming ``bloated'' in that it assumes a radius
substantially larger than it would have if it were an isolated main
sequence star of the same mass.  This bloating can drive mass exchange
between the system components even though a normal main sequence star
of a mass equal to that of the secondary in \exo\ would have a radius
substantially less than the Roche lobe radius and thus would not
transfer substantial amounts of mass to the primary. Indeed,
\citeauthor{hnr01} showed that this effect is greatest in the region of
parameter space around \Porb $\sim 3-4$ hours and $M_2 \sim 0.3$
\Msun, very close to the expected \exo\ system parameters.
Furthermore, if the system is rotating synchronously then the face of
the secondary's surface oriented toward the primary is heated by
X-rays generated by the accretion onto the compact object.  This
heating of the secondary's atmosphere may also lead to substantial
mass transfer without the secondary being in contact with its Roche
lobe (\citealt{rste89}; \citealt{hr91}).

\subsection{The Source of the Intrinsic Period Jitter}

The time scale for both the circularization and synchronization of the
\exo\ orbit is believed to be short compared to the binary
evolutionary time scales (e.g. \citealt{zah77}). However, as
\citet{cha63} has shown for a binary system consisting of two
incompressible fluid masses in synchronous rotation, if the
secondary's rotation rate is fast enough, significant distortions of
the figure of the secondary can occur that are non-axisymmetric.
\citeauthor{cha63} showed that as the value of 
\begin{equation}
\alpha = \frac{\Omega^2}{\pi G \bar{\rho}},
\end{equation}
where $\bar{\rho}$ is the mean density of the secondary star and
$\Omega$ is the angular frequency of the orbital motion, increases for
a given mass ratio the equilibrium figure of the secondary becomes
more extended along the line of centers under the influence of the 
tidal distortions induced by the primary.  
If we assume $M_1 = 1.4 \Msun$ and use \Porb\ to
determine $\Omega$ then $\alpha \sim 0.13$ for any value of the
secondary mass. Assuming $M_2 = 0.35 \Msun$ then from Table \ref{tbl-1} of
\citet{cha63} we find the secondary will be flattened by as much as
20\% at the poles and the equatorial radius will be reduced in the
direction perpendicular to the line of centers by as much as
10\%. Thus, the figure of the secondary is distorted by strong
departures from sphericity.  Furthermore, \citeauthor{cha63} shows
that the effects of gas compressibility will be to enhance the
instability in the secondary component's structure and this could lead
to azimuthal instability modes being excited in the body of the
secondary star. If azimuthal modes of oscillation are excited in the
secondary star this can led to episodic periods of rapid mass transfer
followed by periods of relatively quiescent mass transfer.  Such
oscillation modes need not rotate synchronously with the binary
system, however, and this could result in the system occulting edge as
observed from the earth significantly varying in distance from the
system line of centers and thus causing us to observe seemingly
non-systematic variations in the mid-eclipse times.  Furthermore, the
line of sight as we observe eclipses in \exo\ passes through the
terminator on the secondary star's surface as seen from the X-ray
production region.  The terminator region is likely to be a region of
significant disruption of the secondary's atmosphere, perhaps enough
to also cause variations in the eclipse width as observed at the Earth.

Second, we have previously suggested (Paper II) that the star-spot
cycle in the companion star can lead to changes in the quadrupole moment
of the non-compact mass donating star.  Angular momentum transfer will
then cause variations in the orbital period apparently tied to some
un-seen process occurring within the system.  Transfer of angular
momentum from the spin of the secondary to orbital motion will result
in the secondary star being spun down.  If the angular spin rate of
the secondary is given by $\omega_2$ then we can write the change in
$\omega_2$ corresponding to a change \Porb\ as
\begin{equation}
\Delta\omega_2 = \frac{1}{3} \frac{\Jorb}{I_2} \frac{\Delta\Porb}{\Porb}
\end{equation}
where $I_2 = 2 M_2 R_2^2/5$ is the moment of inertia of the secondary
and $R_2$ is its radius.  If the secondary is approximately tidally
locked and we assume that $R_2$ is given by the Roche lobe radius for
the secondary, using the stellar parameters given above this
corresponds to a change in spin frequency of $\Delta\omega_2 = 2.4
\times 10^{-9}$ Hz, or a fractional change of $5.6 \times 10^{-6}$ in
the spin frequency of the secondary.

However, if the secondary is significantly distorted from spherical by
its tidal interaction with the primary then its moment of inertia will
not be as simple as we assume above: $I_2$ will vary depending on
which axis one considers.  As mentioned above the X-rays from the mass
accretion will asymmetrically heat the secondary's atmosphere.
Convective currents in the secondary will try to redistribute some of
the deposited heat energy around the star.  These convective currents
will redistribute angular momentum in the secondary away from the
distribution that obtains for synchronous rotation.  Enhanced tidal
dissipation of energy associated with non-synchronous rotation will
then be a continuous process for the secondary star.  Also, due to the
presence of the shadowing by the accretion disk around the compact
object the secondary's atmospheric X-ray heating will vary strongly
with latitude. As convection redistributes energy in the outer layers
of the secondary this could give rise to differential
rotation. Differential rotation of the secondary would increase the
need for tidal dissipation of rotational energy in order to maintain
synchronous system rotation.  Furthermore, as large scale convective
currents occur in the outer layers of the secondary star the moments
of inertia for each axis of the secondary will vary by small amounts
as mass is redistributed in a time-dependent fashion.  
Changes in one of the principle axes monents of inertia could
result in the secondary star in \exo\ ``wobbling'' as it rotates in a
manner similar to the ``Chandler wobble'' that occurs in the earth's
polar motion as a result of azimuthal modes in the earth's interior
\citep[see][]{bp93}.  Under these circumstances the \OMC\ diagram for
this system will never settle down into a smooth variation.

If the mass distribution in the donor is sufficiently disrupted so
that its symmetry about the line of centers is affected then small
changes in the apparent orbit period must be made in order to conserve
angular momentum.  If eclipses were purely geometric, then eclipse
transitions are caused by the limb of the companion star.  Changes in
the atmosphere of the companion star which is the occulting edge for
the sharp eclipse transitions, might account for the variable eclipse
profiles (originally suggested by \citealt{psvc91}).  Since the
intrinsic jitter is only of the order $\sim 0.1$ s per eclipse any
departures from a steady shape on the part of the secondary star need
not be large. 
Assuming that $M_2 = 0.40 \Msun$ and the star's radius approaching its
Roche lobe boundary, departures from spherical 
symmetry of order $\sim 40$ km will cause changes of $\sim 0.1$ s 
in the ingress or egress times. 
If the companion star changed shape or size in a manner
which was symmetric about the line between the stars, then one would
expect a correlation between the durations of ingress and egress.  No
such correlation is observed (Figure \ref{fig-durationcorr}).  If the companion star
changed shape or size in a manner which was asymmetric about the same
line and the figure of this shape change moves around the secondary
with the same period as the rotation period of the star, then the
observed eclipse center would move away from the true eclipse center
and the duration of the eclipse would change.  However, no correlation
is observed between mid-eclipse timing residuals and eclipse duration.
On the other hand, if a non-axisymmetric distortion of the secondary's
figure rotates with a period different from the secondary's rotation
period then we might observe eclipse duration variations as we observe
in \exo.  Such variations need not be in phase with the binary period
if they result from instability modes that do not rotate around the
center of the secondary in phase with the orbital period.

\section{Conclusions}

We have presented a three-fold increase in the total number of eclipse
observations available for analysis of the orbital period from \exo.
These eclipse observations take advantage of the sub-second timing
accuracy on the RXTE/PCA and ARGOS/USA experiments to determine
mid-eclipse times to approximately $0.5$ second accuracy. A linear
ephemeris ($\Porbdot = 0$) and a quadratic ephemeris (non-zero
\Porbdot) both give unacceptable fits to the eclipse center timing
data.  A maximum likelihood estimation model for the eclipse center
timings that includes measurement error, intrinsic period jitter, and
orbital period evolution is consistent with the data. We conclude
that the true orbital period in \exo\ has increased by approximately 8 ms
over the past 16 years. When we compare our results for \Porbdot\ to those from
theoretical models of LMXB evolution we find that the model \Porbdot's
have a different sign from our measured \Porbdot\ for the \exo\ system. 

The intrinsic period jitter, which has a magnitude of about 0.1 s per
orbital cycle, causes the mid-eclipse times to do a random walk about
the true solution.  The atmosphere of the secondary star appears to
not be a stable occulting edge on which to base eclipse center timings
and thus its usage as an accurate fiducial marker of orbit evolution
must now be questioned. 
Both the unexpected orbital evolution and the intrinsic period jitter may
result from instabilities in the secondary star that are sufficient to
slightly modify the system angular momentum distribution.  Thus, the
problem presented by the 8 ms change in the orbital period of \exo\
may be related to the problem of the origin of the intrinsic jitter in
the mid-eclipse timings. We have suggested a number of possible
enhancements to theoretical models of LMXB systems that may help solve
both the problem of the unexpected positive orbit period derivative
and the origin of the intrinsic jitter.

\begin{acknowledgements}

We thank Peter Becker, Richard Durisen, Lev Titarchuk, Steven Howell,
and Alan Smale for useful discussions.  We thank David Livingston for
his help in analysis of RXTE data.  We thank an anonymous
referee for his helpful remarks. This work was partially supported
by NASA and by the Office of Naval Research.

\end{acknowledgements}


\clearpage

\begin{figure}
\centerline{\includegraphics[height=6.0in,angle=270.0]{f6.ps}}
\caption{The duration of eclipse transitions (ingress and egress) for \exo\ as
measured by USA and RXTE as a function of observation date for
observations during 1996$-$2000. The circles show the durations of
ingress and the squares show the durations of egress. The duration of
ingress and egress are significantly variable, even for consecutive
eclipses.
\label{fig-transitions}
}
\end{figure}

\begin{figure}
\centerline{\includegraphics[width=6.0in]{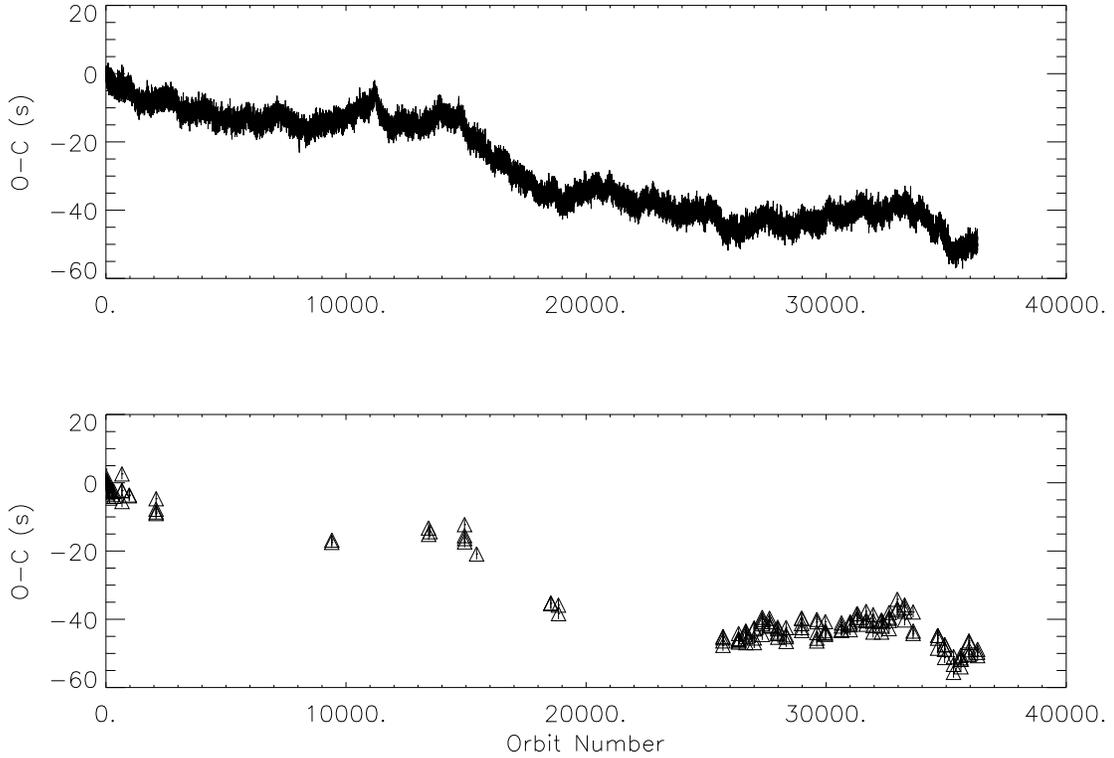}}
\caption{Example output from Monte Carlo simulation of a random
process following Equation~\ref{eqn-randomwalk} with no true orbital
period derivative.  The parameters used are those derived from the
actual data ($\sigma_\epsilon = 0.105$ s and $\sigma_e = 1.62$ s).
The top frame shows the simulated mid-eclipse times for every eclipse,
while the bottom frame shows the same data but sampled using the
sampling function of the actual observational data.  
Notice how simple fitting of short spans of data can easily 
detect spurious positive or negative period derivatives even 
though none actually exists in this case.
\label{fig-monte}
}
\end{figure}

\begin{figure}
\centerline{\includegraphics[height=6.0in,angle=270.0]{f8.ps}}
\caption{The possible values of the period derivative \Porbdot\ as a function
of the secondary mass $M_2$.  The curved solid lines represent the
limits placed on the accretion rate by \citet[$\dot M_1 = 0.4 - 2.2 \,
\times 10^{17} {\rm g} \,\, {\rm s}^{-1}$]{ghpw86} in their analysis
of X-ray bursts from \exo.  We assume a 1.4 \Msun\ neutron star with
radius 10 km.  The vertical dashed lines show the best fit secondary
mass range derived by \citet{pwgg86}: $M_2 = 0.08\, -\, 0.45 \Msun$.
The horizontal dot-dashed line represents $\Porbdot = 1.9 \times
10^{-11}$.
\label{fig-pdot}
}
\end{figure}

\begin{figure}
\centerline{\includegraphics[height=6.0in,angle=270.0]{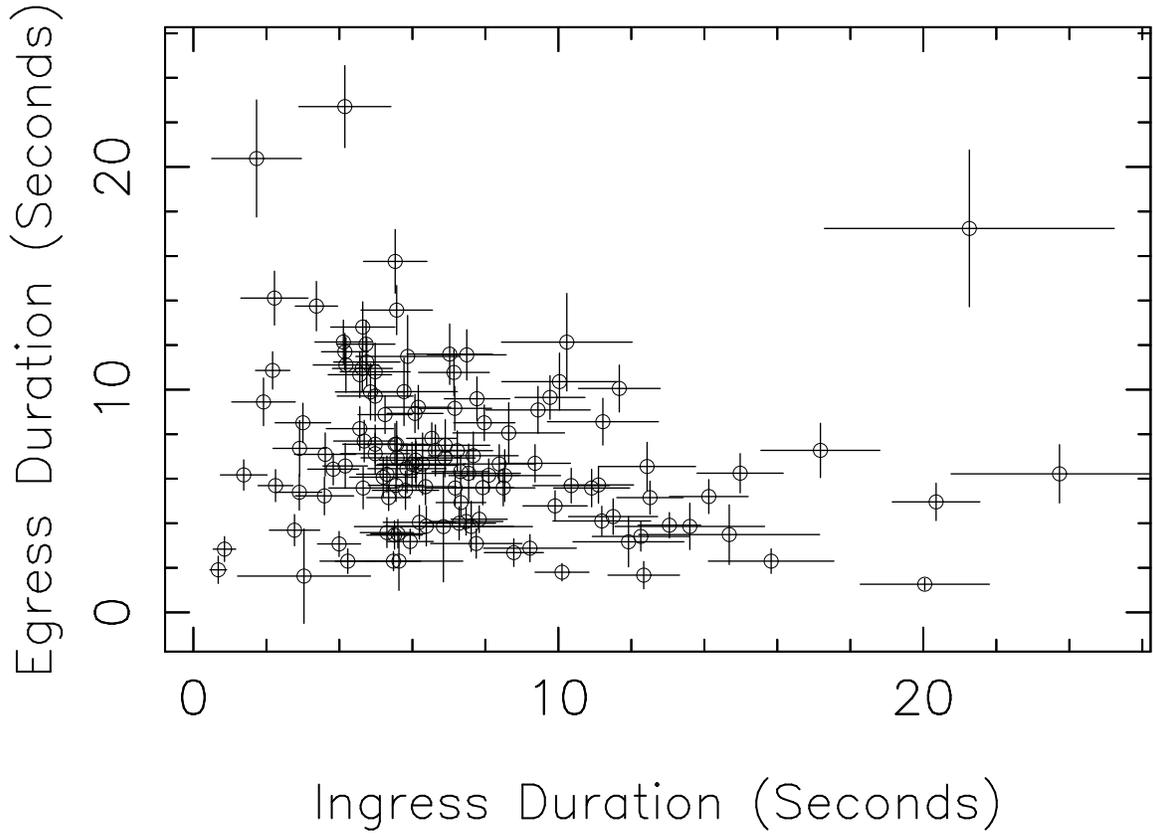}}
\caption{The duration of eclipse egress plotted against the duration of eclipse
ingress for \exo\ as measured by USA and RXTE for observations 
made during 1996$-$2000.
\label{fig-durationcorr}
}
\end{figure}

\clearpage 
%
%


\begin{deluxetable}{cccc}
\tablecolumns{4}
\tablecaption{\bf{RXTE Timing of Full \exo\ Eclipses}\label{tbl-1}}
\tablewidth{380pt}
\tablehead{\colhead{RXTE ObsID} &\colhead{Cycle Number} &\colhead{Mid-Eclipse Time} &\colhead{Timing Uncertainty}\\
  &  & (JD;TDB) & (s) }
\startdata
10108-01-01      &25702 	 &2450206.87478130	  &0.20\\[0.008in]
10108-01-02      &25703 	 &2450207.03409715	  &0.11\\[0.008in]
10108-01-03      &25704 	 &2450207.19343477	  &0.32\\[0.008in]
10108-01-04      &25705 	 &2450207.35277331	  &0.26\\[0.008in]
10108-01-05      &25706 	 &2450207.51211208	  &0.27\\[0.008in]
10108-01-06      &26353 	 &2450310.60371301	  &0.67\\[0.008in]
10108-01-07      &26354 	 &2450310.76305190	  &0.32\\[0.008in]
10108-01-08   	 &26355 	 &2450310.92236498	  &0.43\\[0.008in]
10108-01-09   	 &26356 	 &2450311.08172852	  &0.72\\[0.008in]
10108-01-10   	 &26357 	 &2450311.24103222	  &0.43\\[0.008in]
10108-01-07-01   &26358 	 &2450311.40039275	  &0.32\\[0.008in]
10108-01-11      &26654 	 &2450358.56446486	  &0.22\\[0.008in]
10108-01-12      &26655 	 &2450358.72381394	  &0.14{\tablenotemark{a}}\\[0.008in]
10108-01-13      &26656 	 &2450358.88310803	  &0.26\\[0.008in]
10108-01-14      &26658 	 &2450359.20179449	  &0.62\\[0.008in]
10108-01-15      &26659 	 &2450359.36114044	  &0.37\\[0.008in]
10108-01-15-01   &26660 	 &2450359.52047956	  &0.50\\[0.008in]
20069-01-01 	 &26999 	 &2450413.53600028	  &0.29\\[0.008in]
20069-01-02 	 &27000 	 &2450413.69531428	  &0.57\\[0.008in]
20069-01-03 	 &27001 	 &2450413.85464866	  &0.28\\[0.008in]
20069-01-05 	 &27004 	 &2450414.33266926	  &0.86\\[0.008in]
20069-02-01 	 &27336 	 &2450467.23285063	  &0.27\\[0.008in]
20069-02-02 	 &27338 	 &2450467.55152400	  &0.24\\[0.008in]
20069-02-03 	 &27339 	 &2450467.71085711	  &0.43\\[0.008in]
20069-02-04 	 &27340 	 &2450467.87019530	  &0.22\\[0.008in]
20069-02-05 	 &27341 	 &2450468.02953153	  &0.55\\[0.008in]
20069-03-01 	 &27638 	 &2450515.35283512	  &0.28\\[0.008in]
20069-03-02 	 &27639 	 &2450515.51216336	  &0.51\\[0.008in]
20069-03-03 	 &27640 	 &2450515.67153106	  &0.56\\[0.008in]
20069-03-04 	 &27641 	 &2450515.83084634	  &0.21\\[0.008in]
20069-03-05 	 &27642 	 &2450515.99017262	  &0.22\\[0.008in]
20069-04-01 	 &27981 	 &2450570.00571336	  &0.75\\[0.008in]
20069-04-02 	 &27982 	 &2450570.16505560	  &0.64\\[0.008in]
20069-04-03 	 &27983 	 &2450570.32439866	  &0.37\\[0.008in]
20069-04-04 	 &27984 	 &2450570.48373500	  &0.15\\[0.008in]
20069-04-05 	 &27986 	 &2450570.80243130	  &0.54\\[0.008in]
20069-05-01 	 &28331 	 &2450625.77389125	  &0.81\\[0.008in]
20069-05-02 	 &28332 	 &2450625.93326324	  &0.64\\[0.008in]
20069-05-03 	 &28333 	 &2450626.09259345	  &0.52\\[0.008in]
20069-05-04 	 &28334 	 &2450626.25194090	  &0.51\\[0.008in]
20069-06-01 	 &28976 	 &2450728.54683500	  &0.32\\[0.008in]
20069-06-02 	 &28977 	 &2450728.70615086	  &0.48\\[0.008in]
20069-06-03 	 &28978 	 &2450728.86549287	  &0.33\\[0.008in]
20069-06-04 	 &28979 	 &2450729.02484113	  &0.57\\[0.008in]
20069-06-05 	 &28981 	 &2450729.34351035	  &0.43\\[0.008in]
30067-01-01 	 &29610 	 &2450829.56699252	  &0.48\\[0.008in]
30067-01-02 	 &29611 	 &2450829.72632678	  &0.41\\[0.008in]
30067-01-03 	 &29612 	 &2450829.88563558	  &0.47\\[0.008in]
30067-01-04 	 &29613 	 &2450830.04500850	  &0.43\\[0.008in]
30067-01-05 	 &29614 	 &2450830.20431868	  &0.39\\[0.008in]
30067-02-01 	 &29966 	 &2450886.29134495	  &0.78{\tablenotemark{a}}\\[0.008in]
30067-02-02 	 &29967 	 &2450886.45060676	  &0.43\\[0.008in]
30067-02-03 	 &29968 	 &2450886.60996278	  &0.50\\[0.008in]
30067-02-04 	 &29970 	 &2450886.92862933	  &0.56\\[0.008in]
30067-02-05 	 &29971 	 &2450887.08796799	  &0.39\\[0.008in]
30067-03-01 	 &30636 	 &2450993.04760294	  &0.42\\[0.008in]
30067-03-02 	 &30637 	 &2450993.20693616	  &0.67\\[0.008in]
30067-03-03 	 &30638 	 &2450993.36629113	  &0.64\\[0.008in]
30067-03-04 	 &30640 	 &2450993.68493882	  &0.51\\[0.008in]
30067-03-05 	 &30641 	 &2450993.84429576	  &0.32\\[0.008in]
30067-04-01 	 &30997 	 &2451050.56857736	  &0.38\\[0.008in]
30067-04-02 	 &30998 	 &2451050.72791856	  &0.60\\[0.008in]
30067-04-04 	 &31000 	 &2451051.04656532	  &0.52\\[0.008in]
30067-04-05 	 &31001 	 &2451051.20590051	  &0.66\\[0.008in]
30067-05-01 	 &31292 	 &2451097.57324750	  &0.28\\[0.008in]
30067-05-02 	 &31293 	 &2451097.73256139	  &0.23\\[0.008in]
30067-05-03 	 &31294 	 &2451097.89190097	  &0.51\\[0.008in]
30067-05-04 	 &31295 	 &2451098.05123789	  &0.24\\[0.008in]
30067-05-05 	 &31297 	 &2451098.36990468	  &0.34\\[0.008in]
30067-06-01 	 &31665 	 &2451157.00623616	  &0.33\\[0.008in]
30067-06-02 	 &31666 	 &2451157.16555977	  &0.22\\[0.008in]
30067-06-03 	 &31668 	 &2451157.48419924	  &0.66\\[0.008in]
30067-06-05 	 &31670 	 &2451157.80292725	  &0.40\\[0.008in]
40039-01-01 	 &31955 	 &2451203.21418766	  &0.33{\tablenotemark{a}}\\[0.008in]
40039-01-02 	 &31956 	 &2451203.37353234	  &0.25\\[0.008in]
40039-01-04 	 &31959 	 &2451203.85152829	  &0.22\\[0.008in]
40039-01-05 	 &31960 	 &2451204.01086880	  &0.21\\[0.008in]
40039-02-01 	 &32305 	 &2451258.98247053	  &0.74\\[0.008in]
40039-02-02 	 &32306 	 &2451259.14179623	  &0.44\\[0.008in]
40039-02-03 	 &32308 	 &2451259.46045618	  &0.67\\[0.008in]
40039-02-04 	 &32309 	 &2451259.61978384	  &1.02\\[0.008in]
40039-02-05 	 &32310 	 &2451259.77914981	  &0.26\\[0,010in]
40039-03-01 	 &32624 	 &2451309.81119333	  &0.28\\[0.008in]
40039-03-02 	 &32625 	 &2451309.97057192	  &0.59\\[0.008in]
40039-03-03 	 &32627 	 &2451310.28922956	  &0.14\\[0.008in]
40039-03-04 	 &32628 	 &2451310.44857528	  &0.53\\[0.008in]
40039-03-05 	 &32629 	 &2451310.60793234	  &0.53\\[0.008in]
40039-04-01 	 &32964 	 &2451363.98609206	  &0.47\\[0.008in]
40039-04-02 	 &32965 	 &2451364.14542319	  &0.22\\[0.008in]
40039-04-03 	 &32966 	 &2451364.30478697	  &0.49\\[0.008in]
40039-05-02 	 &33266 	 &2451412.10613465	  &0.36\\[0.008in]
40039-05-03 	 &33267 	 &2451412.26549715	  &0.38\\[0.008in]
40039-05-04 	 &33269 	 &2451412.58410236	  &0.38\\[0.008in]
40039-05-05 	 &33270 	 &2451412.74349356	  &0.57\\[0.008in]
40039-06-02 	 &33622 	 &2451468.83038917	  &0.78\\[0.008in]
40039-06-03 	 &33623 	 &2451468.98974218	  &0.65\\[0.008in]
40039-06-04 	 &33625 	 &2451469.30838454	  &0.37\\[0.008in]
50045-01-01 	 &34647 	 &2451632.15164530	  &0.47\\[0.008in]
50045-01-02 	 &34648 	 &2451632.31097053	  &0.64\\[0.008in]
50045-01-03 	 &34649 	 &2451632.47031914	  &0.60\\[0.008in]
50045-01-05 	 &34652 	 &2451632.94831856	  &0.61\\[0.008in]
50045-02-01 	 &34934 	 &2451677.88157713	  &0.39\\[0.008in]
50045-02-02 	 &34935 	 &2451678.04089808	  &0.32\\[0.008in]
50045-02-03 	 &34937 	 &2451678.35956787	  &0.53\\[0.008in]
50045-02-04 	 &34938 	 &2451678.51892597	  &0.30\\[0.008in]
50045-02-05 	 &34939 	 &2451678.67822018	  &0.36\\[0.008in]
50045-03-01 	 &35304 	 &2451736.83654322	  &0.34\\[0.008in]
50045-03-03 	 &35306 	 &2451737.15523593	  &0.27\\[0.008in]
50045-03-04 	 &35308 	 &2451737.47393280	  &0.61\\[0.008in]
50045-04-01 	 &35610 	 &2451785.59391417	  &0.45\\[0.008in]
50045-04-02 	 &35611 	 &2451785.75322366	  &0.36\\[0.008in]
50045-04-04 	 &35613 	 &2451786.07188373	  &0.37\\[0.008in]
50045-04-05 	 &35614 	 &2451786.23121243	  &0.29\\[0.008in]
50045-04-06 	 &35615 	 &2451786.39058917	  &0.48\\[0.008in]
50045-05-01 	 &35948 	 &2451839.45008512	  &0.40\\[0.008in]
50045-05-02 	 &35949 	 &2451839.60939044	  &0.37\\[0.008in]
50045-05-03 	 &35950 	 &2451839.76875745	  &0.39\\[0.008in]
50045-05-04 	 &35951 	 &2451839.92811845	  &0.56\\[0.008in]
50045-05-05 	 &35952 	 &2451840.08739449	  &0.45\\[0.008in]
50045-06-01 	 &36305 	 &2451896.33361440	  &0.28\\[0.008in]
50045-06-03 	 &36307 	 &2451896.65229738	  &0.47\\[0.008in]
50045-06-04 	 &36309 	 &2451896.97099530	  &0.37
\enddata
\tablenotetext{a}{Analysis time binning is 1.0 seconds for this ObsID.}
\end{deluxetable}




\begin{deluxetable}{cccc}
\tablecolumns{4}
\tablecaption{\bf{USA Timing of Full \exo\ Eclipses}\label{tbl-2}}
\tablewidth{380pt}
\tablehead{\colhead{USA ObsID} &\colhead{Cycle Number} &\colhead{Mid-Eclipse Time} &\colhead{Timing Uncertainty}\\
  &  & (JD;TDB) & (s) }
\startdata
D054-091610-093531 &34432 &2451597.89395861  &1.52\\[0.010in]
D105-011351-014146 &34750 &2451648.56333998  &0.54\\[0.010in]
D109-234506-001510 &34781 &2451653.50289171  &0.19\\[0.010in]
D125-104356-094101 &34878 &2451668.95866382  &0.51\\[0.010in]
D130-091628-081414 &34909 &2451673.89812771  &0.61\\[0.010in]
D135-074949-064725 &34940 &2451678.83764611  &0.79{\tablenotemark{a}}\\[0.010in]
D140-062245-064725 &34971 &2451683.77707690  &1.48
\enddata
\tablenotetext{a}{Analysis time binning is 2.0 seconds for this ObsID.}
\end{deluxetable}


\begin{deluxetable}{ccccc}
\tablecolumns{5}
\tablecaption{\bf{Other Satellite Timing Data of Full \exo\ Eclipses}\label{tbl-3}}
\tablewidth{440pt}
\tablehead{\colhead{Satellite} &\colhead{Reference} &\colhead{Cycle Number} &\colhead{Mid-Eclipse Time} &\colhead{Timing Uncertainty}\\
  &	&  & (JD;TDB) & (s) }
\startdata
EXOSAT  &\citealt{psvc91}  &1    &2446111.734512 &1.5\\[0.010in]
EXOSAT  &\citealt{psvc91}  &2    &2446111.893839 &1.5\\[0.010in]
EXOSAT  &\citealt{psvc91}  &3    &2446112.053182 &1.5\\[0.010in]
EXOSAT  &\citealt{psvc91}  &4    &2446112.212520 &1.5\\[0.010in]
EXOSAT  &\citealt{psvc91}  &5    &2446112.371893 &1.5\\[0.010in]
EXOSAT  &\citealt{psvc91}  &25   &2446115.558604 &1.5\\[0.010in]
EXOSAT  &\citealt{psvc91}  &26   &2446115.717969 &1.5\\[0.010in]
EXOSAT	&\citealt{psvc91}  &27	&2446115.877285 &1.5\\[0.010in]
EXOSAT	&\citealt{psvc91}  &28	&2446116.036640 &1.5\\[0.010in]
EXOSAT	&\citealt{psvc91}  &70	&2446122.728826 &3.0\\[0.010in]
EXOSAT	&\citealt{psvc91}  &71	&2446122.888149 &3.0\\[0.010in]
EXOSAT	&\citealt{psvc91}  &72	&2446123.047517 &3.0\\[0.010in]
EXOSAT	&\citealt{psvc91}  &73	&2446123.206835 &3.0\\[0.010in]
EXOSAT	&\citealt{psvc91}  &146	&2446134.838536 &3.0\\[0.010in]
EXOSAT	&\citealt{psvc91}  &147	&2446134.997833 &3.0\\[0.010in]
EXOSAT	&\citealt{psvc91}  &148	&2446135.157194 &3.0\\[0.010in]
EXOSAT	&\citealt{psvc91}  &263	&2446153.481063 &3.0\\[0.010in]
EXOSAT	&\citealt{psvc91}  &264	&2446153.640374 &3.0\\[0.010in]
EXOSAT	&\citealt{psvc91}  &332	&2446164.475332 &3.0\\[0.010in]
EXOSAT	&\citealt{psvc91}  &333	&2446164.634673 &3.0\\[0.010in]
EXOSAT	&\citealt{psvc91}  &666	&2446217.694155 &1.5\\[0.010in]
EXOSAT	&\citealt{psvc91}  &667	&2446217.853483 &1.5\\[0.010in]
EXOSAT	&\citealt{psvc91}  &668	&2446218.012821 &1.5\\[0.010in]
EXOSAT	&\citealt{psvc91}  &669	&2446218.172160 &1.5\\[0.010in]
EXOSAT	&\citealt{psvc91}  &971	&2446266.292183 &1.5\\[0.010in]
EXOSAT	&\citealt{psvc91}  &972	&2446266.451529 &1.5\\[0.010in]
EXOSAT	&\citealt{psvc91}  &2088	&2446444.272463 &3.0\\[0.010in]
EXOSAT	&\citealt{psvc91}  &2089	&2446444.431786 &3.0\\[0.010in]
EXOSAT	&\citealt{psvc91}  &2090	&2446444.591136 &3.0\\[0.010in]
EXOSAT	&\citealt{psvc91}  &2091	&2446444.750433 &3.0\\[0.010in]
GINGA	&\citealt{adn+92} 	 &9406	&2447610.305887 &1.2\\[0.010in]
GINGA	&\citealt{psvc91}  &9411	&2447611.102559 &0.4\\[0.010in]
GINGA	&\citealt{adn+92} 	 &13438 &2448252.755734 &2.5\\[0.010in]
GINGA	&\citealt{adn+92} 	 &13446 &2448254.030482 &1.0\\[0.010in]
GINGA	&\citealt{adn+92} 	 &13496 &2448261.997296 &2.5\\[0.010in]
GINGA	&\citealt{adn+92} 	 &14939 &2448491.921868 &2.6\\[0.010in]
GINGA	&\citealt{adn+92} 	 &14940 &2448492.081202 &2.8\\[0.010in]
GINGA	&\citealt{adn+92} 	 &14941 &2448492.240537 &2.8\\[0.010in]
GINGA	&\citealt{adn+92} 	 &14942 &2448492.399874 &3.4\\[0.010in]
ROSAT	&\citealt{hlwc94}  &15440 &2448571.750106 &5.0\\[0.010in]
ASCA	&\citealt{cadn94}  &18532 &2449064.422435 &1.5\\[0.010in]
ASCA	&\citealt{cadn94}  &18533 &2449064.581793 &1.5\\[0.010in]
ASCA	&\citealt{cadn94}  &18850 &2449115.091891 &1.5\\[0.010in]
ASCA	&\citealt{cadn94}  &18852 &2449115.410559 &1.5
\enddata
\end{deluxetable}



\begin{deluxetable}{rcl}
\tablecolumns{3}
\tabletypesize{\scriptsize}
\tablecaption{\bf{Orbital Ephemerides of \exo}\label{tbl-4}}
\tablewidth{405pt}
\tablehead{\colhead{Parameter} & & \colhead{Value}}
\startdata
\cutinhead{Constant Period Ephemeris: $T_n = T_0 + n \Porb $ RXTE and USA Data Only}
\phm{XXXXXXXXXXX}$T_0$ (MJD/TDB)\phm{X}&$=$&$46111.07418607 \pm 0.0000030$\\[0.10in]
$\Porb$ (day)\phm{X}&$=$&$0.15933781910 \pm 0.00000000010$\\[0.10in]
$\chi^{2}$(dof)\phm{X}&$=$&$ 84.8 ( 127 )$\\[0.10in]
\cutinhead{Quadratic Ephemeris: $T_n = T_0 + n \Porb + {\frac{1}{2}} n^2 \Porb \Porbdot$ RXTE and USA Data Only}
\phm{XXXXXXXXXXX}$T_0$ (MJD/TDB)\phm{X}&$=$&$46111.072596 \pm 0.000034$\\[0.10in]
$\Porb$ (day)\phm{X}&$=$&$0.1593379245 \pm 0.0000000022$\\[0.10in]
$\Porbdot$\phm{X}&$=$&$(-2.17 \pm 0.05) \times 10^{-11}$\\[0.10in]
$\tau_{orb}$ (yr)\phm{X}&$=$&$2.0 \times 10^{7}$\\[0.10in]
$\chi^{2}$(dof)\phm{X}&$=$&$68.0 ( 126 )$\\[0.10in]
\cutinhead{Quadratic Ephemeris: $T_n = T_0 + n \Porb + {\frac{1}{2}} n^2 \Porb \Porbdot$ All Data}
\phm{XXXXXXXXXXX}$T_0$ (MJD/TDB)\phm{X}&$=$&$46111.0750313 \pm 0.0000037$\\[0.10in]
$\Porb$ (day)\phm{X}&$=$&$0.15933776231 \pm 0.00000000027$\\[0.10in]
$\Porbdot$\phm{X}&$=$&$(1.181 \pm 0.007) \times 10^{-11}$\\[0.10in]
$\tau_{orb}$ (yr)\phm{X}&$=$&$3.7 \times 10^{7}$\\[0.10in]
$\chi^{2}$(dof)\phm{X}&$=$&$121.1 ( 170 )$\\[0.10in]
\cutinhead{Broken Constant Period Ephemeris:
$ T_n  = \left\{ 
\begin{array}{ll}
T_0 + n \Porb_{,0} & \mbox{if $n < n_{break}$} \\
T_0 + n_{break} \Porb_{,0} + ( n - n_{break} ) \Porb_{,1} & \mbox{if $n \geq n_{break}$} 
\end{array} \right. $ All Data}
\phm{XXXXXXXXXXX}$T_0$ (MJD/TDB)\phm{X}&$=$&$46111.0752008 \pm 0.0000042$\\[0.10in]
$\Porb_{,0}$ (day)&$=$&$0.15933772839 \pm 0.00000000066$\\[0.10in]
$n_{break}$ (cycle)\phm{X}&$=$&$11260.8 \pm 62.2$\\[0.10in]
$\Porb_{,1}$ (day)&$=$&$0.159337819444 \pm 0.000000000098$\\[0.10in]
$\chi^{2}$(dof)\phm{X}&$=$&$67.3 ( 169 )$
\enddata

\end{deluxetable}


\begin{deluxetable}{cccc}
\tablecolumns{4}
\tabletypesize{\scriptsize}
\tablecaption{\bf{Maximum Likelihood Estimators for \exo\ Orbital Ephemeris}\label{tbl-5}}
\tablewidth{310pt}
\tablehead{\colhead{Parameter} &\colhead{Full Model} &\colhead{NO Period Change} &\colhead{NO Intrinsic Noise}}
\startdata
\cutinhead{All Data}
Log Likelihood &1576.8254 &1571.7447 &1372.9745 \\[0.10in]
\Porb (day) &$0.1593377355$ &$0.1593377897$ &$0.1593377395$ \\[0.10in]
$\Delta$ (day) &$2.99 \times 10^{-12}$& ... &$3.02 \times 10^{-12}$ \\[0.10in]
q &$0.009087$&$0.01312$&$ ... $ \\[0.10in]
$\sigma_e^2$ (${\rm s}^2$) &$2.34$ &$2.29$ &$49.4$ \\[0.10in]
\Porbdot &$1.87 \times 10^{-11}$& ... &$1.89 \times 10^{-11}$ \\[0.10in]
$\tau_{orb}$ (yr) &$2.3 \times 10^{7}$& ... &$2.3 \times 10^{7}$ \\[0.10in]
$\sigma_e$ (s) &$1.53$ &$1.51$ &$7.03$ \\[0.10in]
$\sigma_{\epsilon}$ (s) &$0.146$ &$0.173$ & ... \\[0.10in]
\cutinhead{RXTE and USA Data Only}
Log Likelihood &1179.7571 &1178.3791 &1141.2872 \\[0.10in]
\Porb (day) &$0.1593380321$ &$0.1593378164$ &$0.1593379030$ \\[0.10in]
$\Delta$ (day) &$-6.96 \times 10^{-12}$& ... &$-2.71 \times 10^{-12}$ \\[0.10in]
q &$0.004341$&$0.005316$&$ ... $ \\[0.10in]
$\sigma_e^2$ (${\rm s}^2$) &$2.66$ &$2.62$ &$7.58$ \\[0.10in]
\Porbdot &$-4.37 \times 10^{-11}$& ... &$-1.70 \times 10^{-11}$ \\[0.10in]
$\tau_{orb}$ (yr) &$1.0 \times 10^{7}$& ... &$2.6 \times 10^{7}$ \\[0.10in]
$\sigma_e$ (s) &$1.63$ &$1.62$ &$2.75$ \\[0.10in]
$\sigma_{\epsilon}$ (s) &$0.107$ &$0.118$ & ... 
\enddata

\end{deluxetable}

\end{document}